\documentclass[aps,showpacs,amsmath,amssymb,twocolumn]{revtex4}
\usepackage{epsfig,bm,dcolumn}
\begin{document}

\title{Preliminary study of the $\bar{K}N$ interaction in a chiral constituent quark model}

\author{F. Huang$^{1}$}
\author{W.L. Wang$^{2}$}
\author{Z.Y. Zhang$^3$}
\author{Y.W. Yu$^3$}
\affiliation{\small
$^1$CCAST (World Laboratory), P.O. Box 8730, Beijing 100080, China \\
$^2$Institute of Particle Physics, Huazhong Normal University, Wuhan 430079, China \\
$^3$Institute of High Energy Physics, P.O. Box 918-4, Beijing
100049, China}

\begin{abstract}

A preliminary investigation of $\bar{K}N$ interaction is performed
within a chiral constituent quark model by solving the resonating
group method (RGM) equation. The model parameters are taken from our
previous work, which gave a satisfactory description of the $S$-,
$P$-, $D$-, $F$-wave $KN$ scattering phase shifts. The
channel-coupling between $\bar{K}N$, $\pi\Lambda$ and $\pi\Sigma$ is
considered, and the scattering phase shifts as well as the
bound-state problem of $\bar{K}N$ are dynamically studied. The
results show that the $S$-wave $\bar{K}N$ interaction in the isospin
$I=0$ channel is attractive, and in the extended chiral SU(3) quark
model such an attraction can make for a $\bar{K}N$ bound state,
which appears as a $\pi\Sigma$ resonance in the coupled-channel
calculation, while the chiral SU(3) quark model cannot accommodate
the existence of a $\bar{K}N$ bound state. It seems that the vector
meson exchanges are necessary to be introduced in the quark-quark
interactions if one tries to explain the $\Lambda(1405)$ as a
$\bar{K}N$ bound state or a $\pi\Sigma$-$\bar{K}N$ resonance state.

\end{abstract}


\maketitle

Nowadays, due to the complexity of the non-perturbative QCD effect,
people still need QCD-inspired models to be a bridge connecting the
QCD fundamental theory and the experimental observables. Among these
phenomenological models, the chiral SU(3) quark model and the
extended chiral SU(3) quark model have been quite successful in
reproducing the energies of the baryon ground states, the binding
energy of the deuteron, the nucleon-nucleon ($NN$) and kaon-nucleon
($KN$) scattering phase shifts, and the hyperon-nucleon ($YN$) cross
sections \cite{zyzhang97,lrdai03,fhuang04kn,fhuang05kne}. In the
original chiral SU(3) quark model, the quark-quark interaction
contains confinement, one gluon exchange (OGE) and boson exchanges
stemming from scalar and pseudoscalar nonets, and meanwhile, it is
found that the short range quark-quark interaction is provided by
OGE and quark exchange effects. In the extended chiral SU(3) quark
model, the vector meson exchanges are included and consequently the
OGE is largely reduced, thus the short range quark-quark interaction
is now dominantly provided by vector-meson exchange and quark
exchange effects.

As a matter of fact, it is still a controversial problem for
low-energy hadron physics whether gluon or Goldstone boson is the
proper effective degree of freedom besides the constituent quark.
Glozman and Riska proposed that the Goldstone boson is the only
other proper effective degree of freedom \cite{glozman96}. But Isgur
insisted that the OGE governs the baryon structure \cite{isgur00}.
Anyway, it is still a challenging problem in the low-energy hadron
physics whether OGE or vector-meson exchange is the right mechanism
or both of them are important for describing the short-range
quark-quark interaction.

In the past few years, by use of the resonating group method (RGM),
we have dynamically studied the $NN$ and $KN$ scattering phase
shifts, and the $\Omega\Omega$, $N\bar\Omega$, $\Delta K$, $\Lambda
K$ and $\Sigma K$ interactions within both the chiral SU(3) quark
model and the extended chiral SU(3) quark model
\cite{zyzhang97,lrdai03,fhuang04kn,fhuang05kne,lrdai06,dzhang07,fhuang05lksk}.
It was found that though the mechanisms of the short range
quark-quark interaction are quite different, these two models give
quite similar results for all those hadron-hadron systems. To get
some more definite information about the short range quark-quark
interaction mechanisms, it seems interesting and necessary to
investigate some special systems where the chiral SU(3) quark model
and the extended chiral SU(3) quark model give different
contributions.

One notices that the $KN$ and $\bar{K}N$ are interesting cases,
since there is a close connection of the vector-meson exchanges
between the $KN$ and $\bar{K}N$ interactions due to $G$-parity
transition on hadron level. Specially, the $\omega$ exchange is
repulsive for $KN$ while attractive for $\bar{K}N$, because of the
negative $G$ parity of the $\omega$ meson. Our previous work
\cite{fhuang05kne} has already shown that for the $KN$ system the
scattering phase shifts from the chiral SU(3) quark model and the
extended chiral SU(3) quark model are quite similar, which means
that in the $KN$ system the repulsion of OGE can be substituted by
that of vector meson exchange. Thus it is naturally expected that
for the $\bar{K}N$ system the results from the chiral SU(3) quark
model and the extended chiral SU(3) quark model must be quite
different since in the $\bar{K}N$ system the OGE vanishes while the
vector-meson exchange is attractive.

In this work, we perform a preliminary study of the $\bar{K}N$
interaction in both the chiral SU(3) quark model and the extended
chiral SU(3) quark model. Let's first briefly review the models (the
detailed formula of which can be found in Ref. \cite{fhuang05kne}).
The total Hamiltonian of baryon-meson systems can be written as
\begin{equation}
H=\sum_{i=1}^{5}T_{i}-T_{G}+\sum_{i<j=1}^{4}V_{ij}+\sum_{i=1}^{4}V_{i\bar
5},
\end{equation}
where $T_G$ is the kinetic energy operator for the center-of-mass
motion, and $V_{ij}$ and $V_{i\bar 5}$ represent the quark-quark and
quark-antiquark interactions, respectively,
\begin{equation}
V_{ij}= V^{\rm OGE}_{ij} + V^{\rm conf}_{ij} + V^{\rm ch}_{ij},
\end{equation}
where $V_{ij}^{\rm OGE}$ is the OGE interaction, $V_{ij}^{\rm conf}$
is the confinement potential, and $V^{\rm ch}_{ij}$ is the chiral
fields induced effective quark-quark potential. In the chiral SU(3)
quark model, $V^{\rm ch}_{ij}$ includes the scalar boson exchanges
and the pseudoscalar boson exchanges,
\begin{eqnarray}
V^{\rm ch}_{ij} = \sum_{a=0}^8 V_{\sigma_a}({\bm
r}_{ij})+\sum_{a=0}^8 V_{\pi_a}({\bm r}_{ij}),
\end{eqnarray}
and in the extended chiral SU(3) quark model, the vector boson
exchanges are also included,
\begin{eqnarray}
V^{\rm ch}_{ij} = \sum_{a=0}^8 V_{\sigma_a}({\bm
r}_{ij})+\sum_{a=0}^8 V_{\pi_a}({\bm r}_{ij})+\sum_{a=0}^8
V_{\rho_a}({\bm r}_{ij}).
\end{eqnarray}
Here $\sigma_{0},...,\sigma_{8}$ are the scalar nonet fields,
$\pi_{0},..,\pi_{8}$ the pseudoscalar nonet fields, and
$\rho_{0},..,\rho_{8}$ the vector nonet fields. $V_{i \bar 5}$ in
Eq. (1) represents the quark-antiquark interaction,
\begin{equation}
V_{i\bar 5}=V_{i\bar 5}^{\rm conf}+V_{i\bar 5}^{\rm OGE}+V_{i\bar
5}^{\rm ch},
\end{equation}
with
\begin{eqnarray}
V_{i\bar{5}}^{\rm ch}=\sum_{j}(-1)^{G_j}V_{i5}^{{\rm ch},j}.
\end{eqnarray}
Here $(-1)^{G_j}$ represents the $G$ parity of the $j$th meson. The
expressions of the potentials can be found in the literature
\cite{fhuang04kn,fhuang05lksk,fhuang05kne}.

All the model parameters are taken to be the same as in Ref.
\cite{fhuang05kne}, which gave a satisfactory description of the
$S$-, $P$-, $D$-, $F$-wave $KN$ phase shifts. Here we briefly give
the procedure for the parameter determination. The three initial
input parameters are taken to be the usual values: $b_u=0.5$ fm for
the chiral SU(3) quark model and $0.45$ fm for the extended chiral
SU(3) quark model, $m_{u(d)}=313$ MeV, and $m_s=470$ MeV. The
coupling constant for scalar and pseudoscalar chiral field coupling,
$g_{\rm ch}$, is fixed by the relation
\begin{eqnarray}
\frac{g^{2}_{\rm ch}}{4\pi} = \left( \frac{3}{5} \right)^{2}
\frac{g^{2}_{NN\pi}}{4\pi} \frac{m^{2}_{u}}{M^{2}_{N}},
\end{eqnarray}
with the empirical value $g^{2}_{NN\pi}/4\pi=13.67$. The coupling
constant for vector coupling of vector-meson field is taken to be
$g_{\rm chv}=2.351$, the same as used in the $NN$ case
\cite{lrdai03}. The OGE coupling constants and the strengths of
confinement are fitted by baryon masses and their stability
conditions. All the parameters are tabulated in Table I.

{\small
\begin{table}[thb]
\caption{\label{para} Model parameters. The meson masses and the
cutoff masses: $m_{\sigma'}=980$ MeV, $m_{\kappa}=1430$ MeV,
$m_{\sigma}=675$ MeV, $m_{\epsilon}=980$ MeV, $m_{\pi}=138$ MeV,
$m_K=495$ MeV, $m_{\eta}=549$ MeV, $m_{\eta'}=957$ MeV,
$m_{\rho}=770$ MeV, $m_{K^*}=892$ MeV, $m_{\omega}=782$ MeV,
$m_{\phi}=1020$ MeV, $\Lambda=1500$ MeV for $\kappa$ and 1100 MeV
for other mesons.}
\begin{center}
\begin{tabular*}{85mm}{@{\extracolsep\fill}lcc}
\hline\hline
  & $\chi$-SU(3) QM & Ex. $\chi$-SU(3) QM  \\
\hline
 $b_u$ (fm)  & 0.5 & 0.45  \\
 $m_u$ (MeV) & 313 & 313  \\
 $m_s$ (MeV) & 470 & 470  \\
 $g_u^2$     & 0.7704 & 0.0748 \\
 $g_s^2$     & 0.5525 & 0.0001 \\
 $g_{\rm ch}$    & 2.621 & 2.621  \\
 $g_{\rm chv}$   &       & 2.351  \\
 $a^c_{uu}$ (MeV/fm$^2$) & 55.7 & 60.3  \\
 $a^c_{us}$ (MeV/fm$^2$) & 72.1 & 98.8  \\
 $a^{c0}_{uu}$ (MeV)  & $-$56.4 & $-$91.8 \\
 $a^{c0}_{us}$ (MeV)  & $-$63.0 & $-$116.8  \\
\hline\hline
\end{tabular*}
\end{center}
\end{table}}

From Table I one can see that in the extended chiral SU(3) quark
model, $g_u^2$ and $g_s^2$ are much smaller than the values in the
chiral SU(3) quark model. This means that when the coupling of
quarks and vector-meson field is considered, the coupling constants
of OGE are greatly reduced by fitting the mass splits of $m_\Delta -
m_N$ and $m_\Sigma - m_\Lambda$. Thus the OGE that plays an
important role of the quark-quark short-range interaction in the
original chiral SU(3) quark model is nearly replaced by the
vector-meson exchange in the extended chiral SU(3) quark model. In
other words, the mechanisms of the quark-quark short-range
interactions in these two models are quite different.

Our previous work \cite{fhuang05kne} has already shown that both
these two sets of parameters can give a satisfactory description of
the $S$-, $P$-, $D$-, $F$-wave $KN$ phase shifts. In this work, we
use the same parameters to perform a preliminary study of the ${\bar
K}N$ interaction by use of the RGM.

\begin{figure}[t!]
\epsfig{file=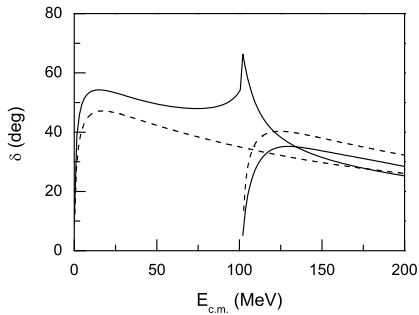,width=5.5cm} \vglue -0.2cm
\caption{\small \label{phase} $S$-wave ${\bar K}N$ and $\pi\Sigma$
($I=0$) phase shifts in the chiral SU(3) quark model. The solid and
dashed lines denote the results of single-channel and
coupled-channels calculations, respectively. The $\bar KN$ phase
shifts start at $E_{\rm c.m.}=102$ MeV.}
\end{figure}

\begin{figure}[b!]
\epsfig{file=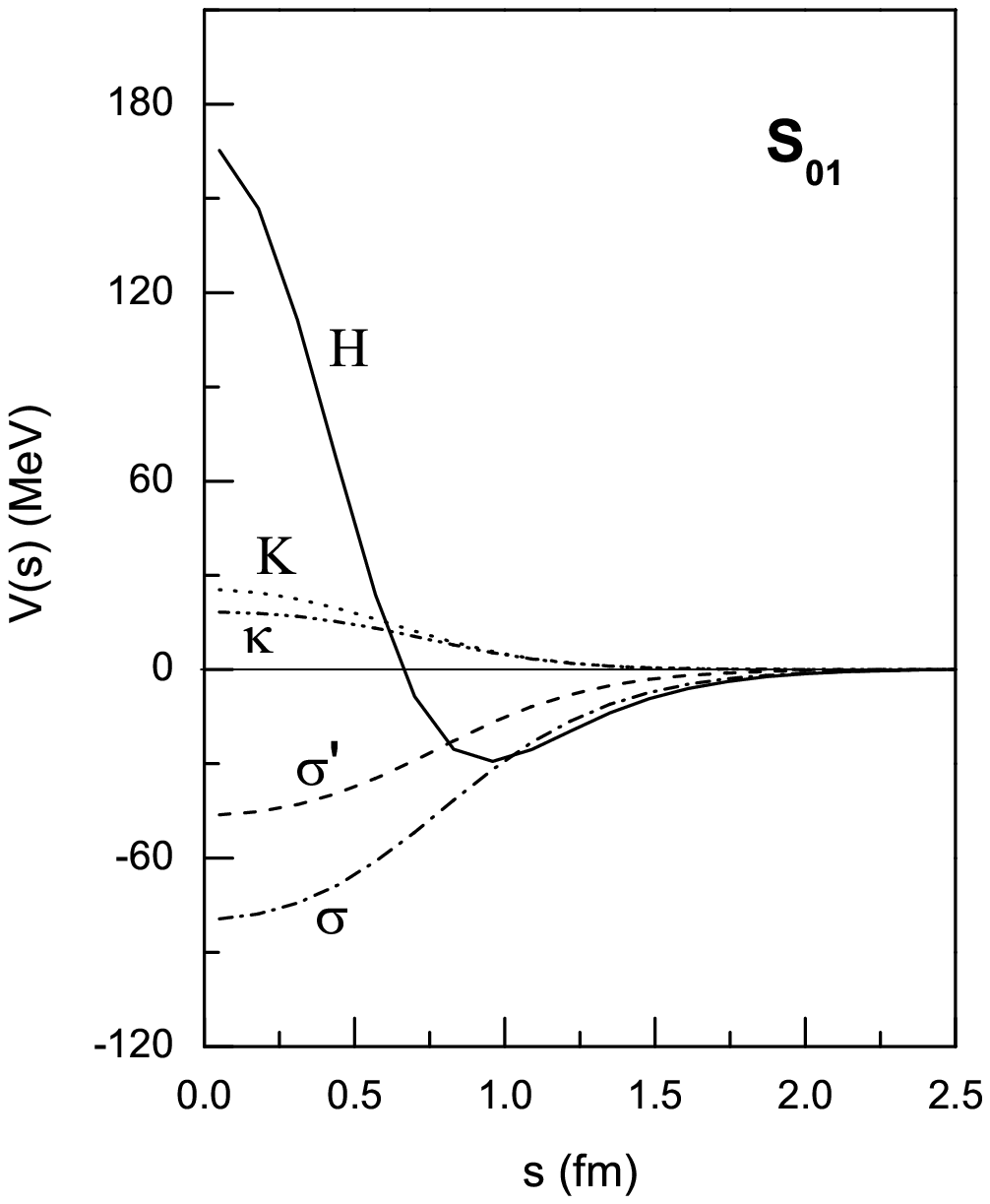,width=5.0cm,height=5.2cm} \vglue -0.2cm
\caption{\small \label{v} GCM matrix elements for the $S$-wave
${\bar K}N$ system ($I=0$) in the chiral SU(3) quark model.}
\end{figure}

Fig. \ref{phase} shows the $S$-wave ${\bar K}N$ and $\pi\Sigma$
($I=0$) scattering phase shifts calculated in the chiral SU(3) quark
model. One sees that the phase shifts of both $\pi\Sigma$ and ${\bar
K}N$ ($I=0$) are positive, which consequently means that their
interactions are attractive, and besides, the effect of
channel-coupling between ${\bar K}N$ and $\pi\Sigma$ is
considerable, which is because of the big off-diagonal matrix
elements of ${\bar K}N-\pi\Sigma$. We further solved the RGM
equation for a bound state problem, and the results show that there
is no $\pi\Sigma$ or ${\bar K}N$ bound state since the strengths of
their attractions are not strong enough. To understand the
contributions of various chiral fields to the ${\bar K}N$
interaction, in Fig. \ref{v} we show the diagonal matrix elements of
the generator coordinate method (GCM) calculation \cite{kwi77} in
the chiral SU(3) quark model, which can describe the interaction
between two clusters ${\bar K}$ and $N$ qualitatively. In Fig.
\ref{v}, $s$ denotes the generator coordinate and $V(s)$ is the
effective interacting potential between the two clusters. One sees
that the attractive interaction of ${\bar K}N$ is dominantly
provided by the $\sigma$ and $\sigma'$ exchanges, but the strengths
of the attractions are not very large, so that a bound state is
unable to be formed. Nevertheless on the hadron level, the ${\bar
K}N$ interaction has been widely investigated by use of the
effective chiral Lagrange method \cite{kaiser95}, and it is found
that the ${\bar K}N$ interaction is attractive enough to form a
bound state. In the chiral SU(3) quark model, if we take the
parameters as the same as in the $KN$ scattering study, we cannot
get a ${\bar K}N$ bound state since the attraction strength is not
strong enough. If we adjust the mass of $\sigma$ to a smaller value,
the ${\bar K}N$ interaction will be much more attractive and it may
be possible to get a ${\bar K}N$ bound state, but then the $KN$
scattering cannot be described simultaneously.

\begin{figure}[t!]
\epsfig{file=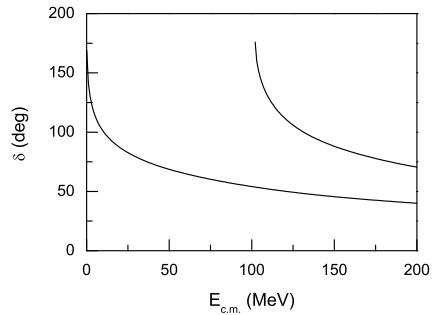,width=5.5cm} \vglue -0.2cm
\caption{\small \label{phase-e0} $S$-wave ${\bar K}N$ and
$\pi\Sigma$ ($I=0$) scattering phase shifts in one-channel
calculation in the extended chiral SU(3) quark model. The phase
shifts for $\bar KN$ scattering start at $E_{\rm c.m.}=102$ MeV.}
\end{figure}

\begin{figure}[b!]
\epsfig{file=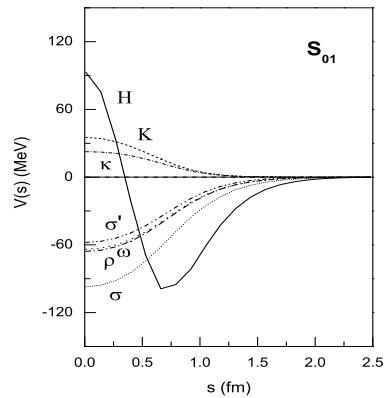,width=5.0cm,height=5.2cm} \vglue -0.2cm
\caption{\small \label{v-e} GCM matrix elements for the $S$-wave
${\bar K}N$ system ($I=0$) in the extended chiral SU(3) quark
model.}
\end{figure}

Fig. \ref{phase-e0} shows the $S$-wave one-channel ${\bar K}N$ and
$\pi\Sigma$ ($I=0$) phase shifts in the extended chiral SU(3) quark
model. The phase shifts indicate that both $\pi\Sigma$ and ${\bar
K}N$ have very strong attractive interactions. Further study shows
that such strong attractive interactions can make for $\pi\Sigma$
and ${\bar K}N$ bound states with the binding energies of about $4$
and $12$ MeV, respectively. Fig. \ref{v-e} presents the diagonal GCM
matrix elements of ${\bar K}N$, which denotes the contributions of
various interaction terms. Comparing with Fig. \ref{v}, one sees
that besides $\sigma'$ and $\sigma$, $\rho$ and $\omega$ also offer
strong attractions, which makes the $\bar KN$ interaction much more
strongly attractive. The $\pi\Sigma$ is also much more attractive
due to the contributions of vector meson exchanges. Such strong
attractions consequently make for the $\pi\Sigma$ and ${\bar K}N$
bound states.

We know from Ref. \cite{fhuang05kne} that for the $KN$ system, both
the chiral SU(3) quark model and the extended chiral SU(3) quark
model give similar contributions, though the short-range interaction
mechanisms are quite different in these two models. This means for
the $KN$ system, the repulsion of OGE can be replaced by that of
vector meson exchange. While for the $\bar KN$ system, OGE totally
vanishes and vector meson exchange contributes strong attractions.
Thus the results from those two models are quite different: neither
$\pi\Sigma$ nor $\bar KN$ is bound in the chiral SU(3) quark model,
while both $\pi\Sigma$ and $\bar KN$ are bound in the extended
chiral SU(3) quark model. This means the attractions stemming from
vector meson exchanges play a dominant role in the forming of the
$\pi\Sigma$ and $\bar KN$ bound states, which is very similar to the
works done at the hadron level with the effective chiral Lagrangian,
where the lowest order chiral Lagrangian for meson-baryon
interaction produces the Weinberg-Tomozawa term, which is nothing
else than the exchange of vector mesons and does indeed produce
attraction in the $\bar KN$ channel.

\begin{figure}[t!]
\epsfig{file=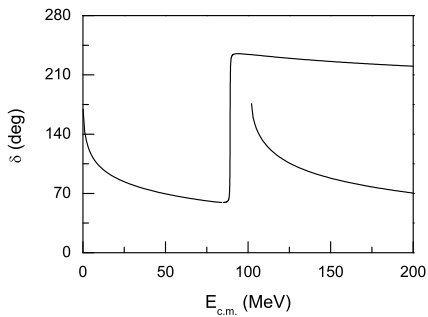,width=5.5cm} \vglue -0.2cm
\caption{\small \label{phase-ec} $S$-wave ${\bar K}N$ and
$\pi\Sigma$ ($I=0$) scattering phase shifts in coupled-channels
calculation in the extended chiral SU(3) quark model. The $\bar KN$
phase shifts start at $E_{\rm c.m.}=102$ MeV.}
\end{figure}

The appearance of the $\pi\Sigma$ and $\bar KN$ bound states in our
model is very interesting, since on the hadron level all models
using effective chiral Lagrangian have already reported two poles
close to the $\Lambda(1405)$: the higher one couples mostly to $\bar
KN$ and the lower one mostly to $\pi\Sigma$ \cite{kaiser95}. We
further perform a $\bar KN$-$\pi\Sigma$ coupled-channels
calculation, and the phase shifts are shown in Fig. \ref{phase-ec}.
From this figure one sees that the $\bar KN$ bound state in the
one-channel calculation appears as a sharp $\pi\Sigma$ resonance
just below the $\bar KN$ threshold in the coupled-channels
investigation. In Ref. \cite{takeuchi07}, Takeuchi and Shimizu
claimed that the coupling of a $\pi\Sigma$ bound state and a
three-quark state will make for a resonance near $\Lambda(1405)$.
Thus it can be naturally expected that in the extended chiral SU(3)
quark model another resonance with dominant $\pi\Sigma$ component
would appear below the $\bar KN$ threshold with the inclusion of the
coupling of a three-quark state to the $\pi\Sigma$-$\bar KN$
channels. If so, of great interest is that one comes to the same
physics even using different approaches, i.e. the chiral quark model
and the effective chiral Lagrangian method. Investigations along
this line are planned for our next-step work.

\begin{figure}[t!]
\epsfig{file=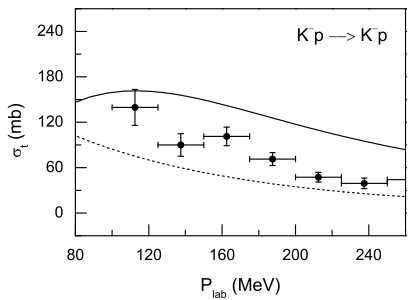,width=5.5cm} \vglue -0.2cm \caption{\small
\label{kp} Cross sections of $K^-p \rightarrow K^-p$. The dashed and
solid lines represent the results from the chiral SU(3) quark model
and the extended chiral SU(3) quark model, respectively. The
experimental data are taken from Ref. \cite{humphrey62}.}
\end{figure}

Fig. \ref{kp} shows the $K^-p \rightarrow K^-p$ cross sections. One
sees that the theoretical results are not far from the experimental
data in both the chiral SU(3) quark model and the extended chiral
SU(3) quark model. Nevertheless they are not good enough, which
implies that the coupling of a genuine three-quark component to
$\bar KN$-$\pi\Sigma$ should be included in order to improve the
description.

In summary, we perform a preliminary study of the $\bar KN$
interaction using the same model, the same parameters and the same
method as in the $KN$ phase shifts study \cite{fhuang05kne}. It is
found that the chiral SU(3) quark model cannot accommodate a ${\bar
K}N$ bound state or a $\pi\Sigma$-${\bar K}N$ resonance state, while
in the extended chiral SU(3) quark model, both $\pi\Sigma$ and $\bar
KN$ are bound and the latter appears as a $\pi\Sigma$ resonance in
the coupled-channels calculation. We conclude that the vector meson
exchanges are necessary to be included in the quark-quark
interaction potentials if one tries to explain the $\Lambda(1405)$
as $\bar{K}N$ bound state or $\pi\Sigma$-$\bar{K}N$ resonance state.
The $\bar{K}N$ cross sections and the effects of the three quark
component will be studied in future work.

This work was supported in part by the National Natural Science
Foundation of China (No. 10475087) and China Postdoctoral Science
Foundation.

\end{document}